\documentclass[aps,prl,twocolumn,showpacs,superscriptaddress,groupedaddress]{revtex4}
\usepackage{graphicx}
\usepackage{dcolumn}
\usepackage{bm}
\usepackage{amssymb}
\usepackage{eft}
\usepackage{amsmath}
\usepackage{verbatim}
\usepackage[svgnames,hyperref]{xcolor}
\usepackage{subfigure}

\begin{document}

\title{Ordering multiple soft gluon emissions using Glauber-SCET}
\author{Jeffrey Robert Forshaw}
\author{Jack Holguin}
\author{Aditya Pathak}
\affiliation{University of Manchester, School of Physics and Astronomy, Manchester, M13 9PL, United Kingdom}
\date{\today}

\begin{abstract}
In QCD, the amplitude for multiple soft gluon emissions has been shown to exhibit a form of coherence, wherein one-loop corrections are rendered IR finite by the transverse momentum of the real emissions. Using a sequence of soft-collinear effective theories (SCET) with Glauber operators, we present a compact and elegant derivation of this property. Our derivation highlights an important role played by the Lipatov vertex. The EFT approach provides increased physical intuition and, for the first time, a clear road map for how this result can be extended to higher loop orders.
\end{abstract}

\maketitle

There is a remarkable property in perturbative QCD, so far proven only at one-loop accuracy, whereby the amplitude for soft gluon emission off a generic hard process can be organized in such a way that the infra-red (IR) divergences associated with the loop integrals can be tamed \cite{Angeles-Martinez:2016dph}. Specifically, the amplitude for multi-gluon emission can be written as a sum over Markovian chains of real emissions with the loop inserted into the chain at all possible places. The associated loop integrals are all rendered IR finite by the transverse momentum of the next emission in the chain. As a result, only the case where the loop integral occurs at the end of the chain is IR divergent.

The original derivation of this result \cite{Angeles-Martinez:2015rna,Angeles-Martinez:2016dph} involved a complicated sum over many Feynman diagrams, which makes the extension to higher orders a formidable task. In this \textit{letter}, we re-derive this result in the framework of soft collinear effective theory (SCET)~\cite{Bauer:2000ew,Bauer:2000yr,Bauer:2001yt,Bauer:2001ct,Bauer:2002nz} with Glauber potential operators~\cite{Rothstein:2016bsq}. The use of SCET drastically simplifies the analysis with far fewer diagrams and a clear physical picture.
In other words, the calculation in SCET is organized in such a way that the appropriate grouping of the QCD diagrams is automatic. This should pave the way to an all-loops calculation in the future.
Our use of SCET is also novel in the sense that we have used it as a tool for a fixed-order calculation in a multi-scale problem.

\subsection{Introduction}

Soft gluon factorization is an important property of QCD (e.g. see \cite{Bassetto:1983mvz,Bern:1999ry,CATANI2000435,Duhr:2013msa,Li:2013lsa,Feige:2014wja}). Accordingly, the amplitude for the emission of $N$ soft gluons off an $n$-parton hard scattering process, $|M_N \big] =| {\cal M} (q_1,\ldots,q_N, p_1,\ldots p_n) \big ]$, can be written as a product of soft gluon emission operators acting on a dressed hard scattering vector:
\begin{align}\label{eq:Ma1}
| M_N \big ] &\simeq (g \mu^\eps)^N
\mb J (q_N) \cdots \mb J (q_1) \big | M_0 \big] \, ,
\end{align}
where $q_i$ is a soft gluon momentum and $p_i$ is a hard parton. The soft momenta have a hierarchy so that $q_i$ is more soft than $q_{i-1}$. The bracket notation indicates vectors in SU(3) colour space. The hard scattering vector and the soft gluon operators have loop expansions~\cite{CATANI2000435}:
\begin{align}\label{eq:JMFact1}
| M_0 ] &= \big| M_0^{(0)}\big ] + \big| M_0^{(1)}\big ] + \ldots
\, \nonumber \\
\text{and~~~~} \mb J({q}) & = \mb J^{(0)} (q) + \mb J^{(1)} (q) + \ldots \, ,
\end{align}
where the superscripts denote the loop order. At tree level,
\begin{align}\label{eq:J0}
\mb J^{(0)} (q_{m+1}) &= \sum_{j = 1}^{n+m} \mb T_j \frac{p_j \cdot \veps}{p_j\cdot q_{m+1}}
= \sum_{j = 1}^{n+m} \mb d_{ij}^{(0)}(q_{m+1}), \, \nonumber \\
\mb d_{ij}^{(0)}(q) &= \mb T_j \Big(\frac{p_j \cdot \veps}{ p_j \cdot q} - \frac{p_i \cdot \veps}{p_i \cdot q}\Big)
\,
\end{align}
for any choice of parton $i$ ($\veps$ is the polarization vector of the emitted gluon) and where $p_{n+m}\equiv q_{m}$. And at one loop,
\begin{align}\label{eq:J1}
\mb J^{(1)} &(q_{m+1}) = \sum_{j = 1}^{n+m} \sum_{k=1}^{n+m} \mb d_{jk}^{(1)}(q_{m+1}), \, \\
\mb d_{ij}^{(1)} (q) &= \frac{\alpha_s}{2\pi} \frac{c_\Gamma}{\eps^2} \mb T_{q} \cdot \mb T_i \Bigg(\frac{ e^{-\im \pi \tilde \delta_{ij}}}{e^{-\im \pi \tilde \delta_{iq}}e^{-\im \pi \tilde \delta_{jq}}} \frac{4\pi \mu^2}{\big({q_{\perp}^{(ij)}}\big)^2}\Bigg)^\eps
\mb d_{ij}^{(0)}(q) \, , \nn
\end{align}
with $c_\Gamma = \frac{\Gamma^3(1-\eps)\Gamma^2(1+\eps)}{\Gamma(1-2\eps)}$.
The corresponding one-loop hard scattering vector is
\begin{align}
| M_0^{(1)} \big] &= \sum_{i=2}^{n} \sum_{j=1}^{i-1} \; \mb I_{ij}(0,\omega_{ij}) \, | M_0^{(0)} \big]~,
\end{align}
where $\omega_{ij} = 2 p_i \cdot p_j$ and the loop correction is given by
\begin{align}
\mb I_{ij}(0,\omega_{ij}) &=
\frac{\alpha_s}{2\pi}
\frac{ c_\Gamma}{\eps^2}
\mb T_i \cdot \mb T_j
\Bigg(e^{\im \pi \tilde \delta_{ij}}\frac{4\pi \mu^2 }{\omega_{ij}}\Bigg)^\eps
\, ,
\end{align}
where $\tilde{\delta}_{ij} = 1$ if partons $i$ and $j$ are either both incoming or both outgoing and $\tilde{\delta}_{ij} = 0$ otherwise. In this way of approaching things, both the soft gluon emission operators and the hard scattering vector are infrared (IR) divergent (i.e. they have $1/\eps$ poles).

Remarkably, at one-loop at least, this result for the amplitude can be rewritten as a chain of real emissions with a loop inserted at any point in the chain. Crucially, all of the loop integrals except for the last one are rendered finite since they are cutoff by the transverse momentum of the next emission in the chain (evaluated in an appropriate frame). Specifically, the one-loop amplitude with $N$ emissions can be written:
\begin{widetext}
\begin{align}\label{eq:AMFS}
\big | M^{(1)}_{\gr N} \big ] &=
(g\mu^{\eps})^{\textcolor{Green}{N}} \bigg( \prod_{k = 1}^{\gr N} \mb J^{(0)}(\textcolor{Green}{q_k})\bigg)
\bigg(
\sum_{\textcolor{Red}{i} = 2}^{n} \sum_{\textcolor{Red}{j} < \textcolor{Red}{i}}
\mathbf{I}^{(\textcolor{Red}{ij})}( \textcolor{Green}{q_{1\perp}^{(\textcolor{Red}{ij})}}, \textcolor{Red}{\sqrt{\omega_{ij}}})
\bigg) \big|M_0^{(0)} \big ]
\\
&+ (g\mu^{\eps})^{\textcolor{Green}{N}} \sum_{m= 1}^{\textcolor{Green}{N}}
\bigg( \prod_{k = {m} + 1}^{\gr N} \mb J^{(0)}(\textcolor{Green}{q_k})\bigg)
\bigg(
\sum_{\textcolor{Red}{i} = 2}^{n + m-1} \sum_{\textcolor{Red}{j} <\textcolor{Red}{i}}
\mb I^{(\textcolor{Red}{ij})} ( \textcolor{Green}{q_{m+1\perp}^{(\textcolor{Red}{ij})}},
\bl{q^{(\textcolor{Red}{ij})}_{m\perp}})\,
\bigg)
\bigg( \prod_{\ell = 1}^{m} \mb J^{(0)}(\bl{q_\ell})\bigg)
\big | M_0^{(0)} \big ]
\nn
\\
&+ (g\mu^{\eps})^{\textcolor{Green}{N}} \sum_{m = 1}^{\textcolor{Green}{N}}
\bigg( \prod_{k = {m} + 1}^{\gr N} \mb J^{(0)}(\textcolor{Green}{q_k})\bigg)
\bigg(
\sum_{\textcolor{Red}{i}, \textcolor{Red}{j} = 1}^{n+m-1}
\mathbf{I}^{(\textcolor{Blue}{(n+m) }\textcolor{Red}{i})} ( \textcolor{Green}{q_{m+1\perp}^{(\textcolor{Blue}{(n+m)}\textcolor{Red}{i})}},
\textcolor{Blue}{q_{m\perp}^{(\textcolor{Red}{ij})}})
\mb d^{(0)}_{\textcolor{Red}{ij}} (\textcolor{Blue}{q_m}) \bigg)
\bigg( \prod_{\ell = 1}^{m - 1} \mb J^{(0)}(\bl{q_\ell})\bigg)
\big | M^{(0)}_0 \big ]
\nn \, ,
\end{align}
where
\begin{align}\label{eq:Iab}
\mb I_{ij}(a,b) = \frac{\alpha_s }{2\pi} \mb T_i \cdot \mb T_j \frac{c_\Gamma}{\eps^2}
\bigg[
\Big(\frac{4\pi \mu^2}{b^2}\Big)^\eps
\Big(
1 + \im \pi \eps \tilde \delta_{ij} - \eps \ln \frac{\omega_{ij}}{b^2}
\Big)
-
\Big(\frac{4\pi \mu^2}{a^2}\Big)^\eps
\Big(
1 + \im \pi \eps \tilde \delta_{ij} - \eps \ln \frac{\omega_{ij}}{a^2}
\Big)
\bigg] + \mathrm{Re} ~ \mathcal{O}(\epsilon^{0}) \, .
\end{align}
\end{widetext}
The $\perp$ subscript is defined so that $q_\perp^{(ij)}$ is the transverse momentum defined in the $ij$ zero momentum frame. Note that the real part of \eq{AMFS} can be obtained from the imaginary part by analytic continuation:
\begin{align}
\im \pi \eps \tilde \delta_{ij} - \eps \ln \frac{\omega_{ij}}{\mu^2} = \eps \ln \frac{\mu^2}{-\omega_{ij}}.
\end{align}

\begin{figure}[t!]
\includegraphics[width=.45\textwidth]{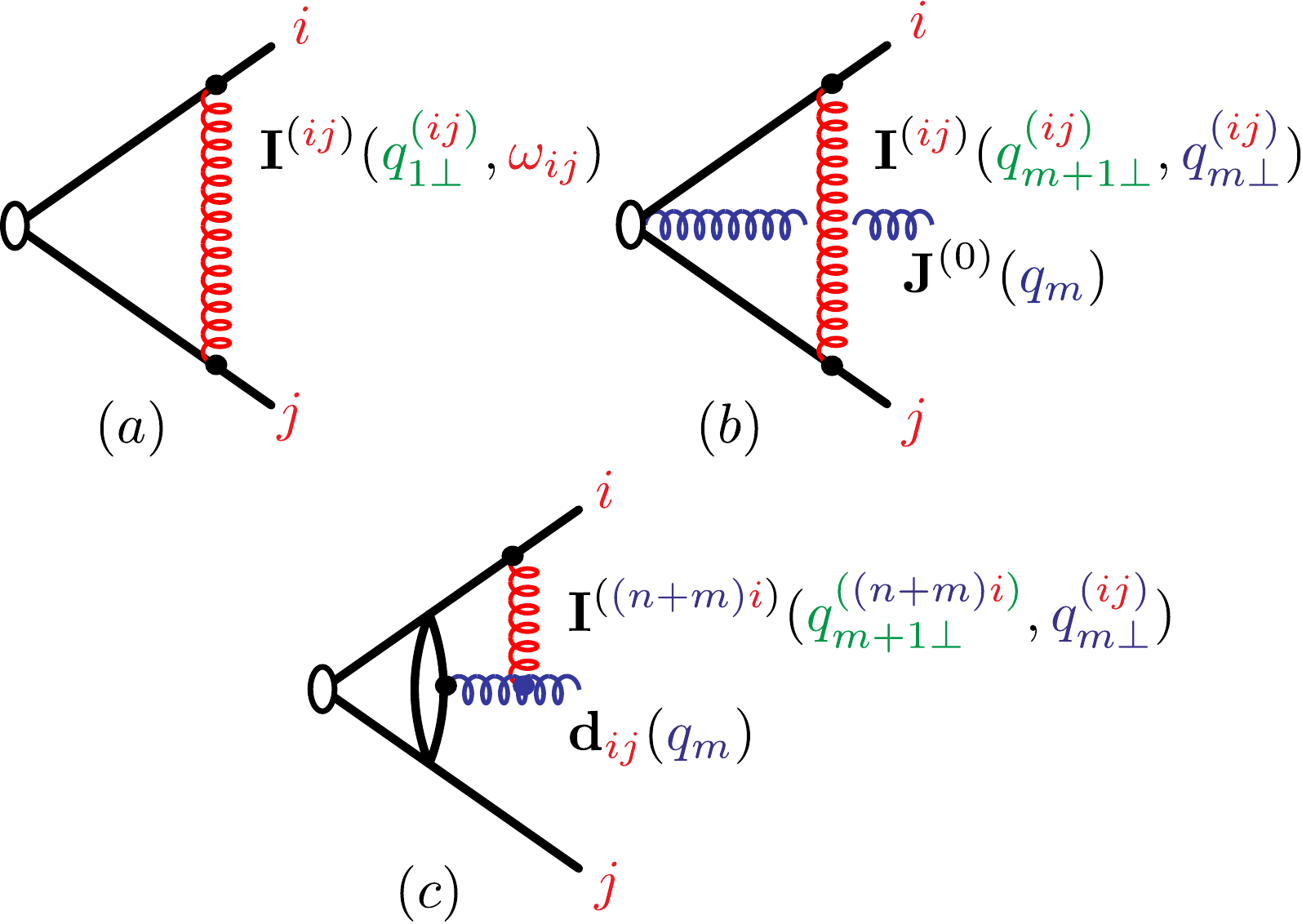}
\caption{Relation of limits of the virtual loop insertions to the adjacent dipole emission momenta in the ordered soft gluon emission result. The graphs (a), (b) and (c) correspond to the three lines in \eq{AMFS}.
\label{fig:AMFS}}
\end{figure}

The result can be understood by relating each of the three lines in \eq{AMFS} with the three diagrams shown in \fig{AMFS} (the color coding in \eq{AMFS} agrees with the colors in \fig{AMFS} and later in \eq{AMFS_SCET}). The loop insertions are given by $\mb I_{ij}(a,b)$ in \eq{Iab}. The first argument represents the lower limit of the loop integral and it is determined by the next soft gluon emission in the chain (this gluon is not shown in the figure). The second argument represents the upper limit and it is determined by the previous soft gluon emission. In the first line, the upper limit is simply the hard scale (there are no prior emission). So long as $a,b > 0$, one can safely take the limit $\eps \to 0$ in \eq{Iab}. The dipole frame in which loop cutoff momenta are evaluated carries special significance. In the first two lines it is just the frame defined by the two partons between which the virtual gluon is exchanged. The third line is interesting for it reveals that the gluon's transverse momentum is to be evaluated in the rest frame of its \textit{parent-dipole}. Because of this, the individual dipole contributions cannot be summed over in this term, preventing the result from being written in terms of $\mb J$ operators.

\begin{figure}[t!]
\centering
\subfigure[{The Glauber exchange graph in EFT$_n$.
\label{fig:all1} }] {\includegraphics[width=0.4\textwidth]{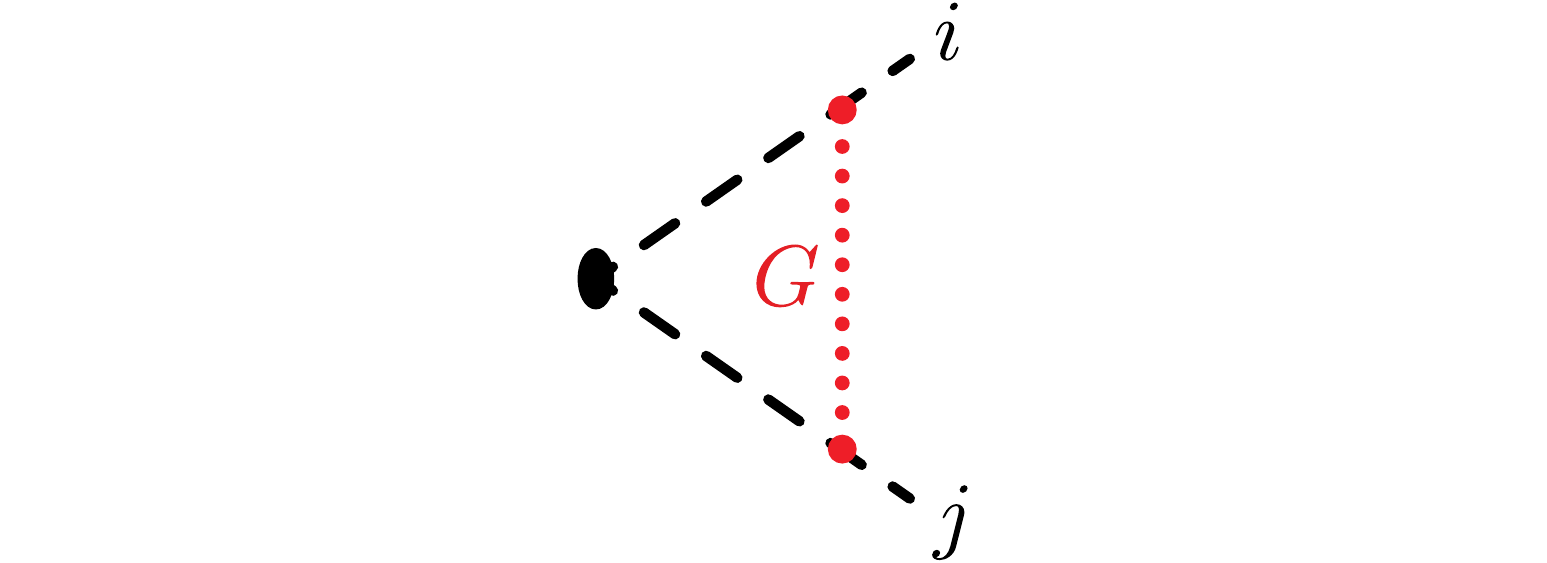}} \\ \vspace{5pt}
\subfigure[{Single soft gluon emission graphs with a Glauber exchange in EFT$_{n}$. The second graph involves the Lipatov vertex.
Additional graphs with permutations of $i$ and $j$ are not shown.
\label{fig:all2} }] {\includegraphics[width=0.4\textwidth]{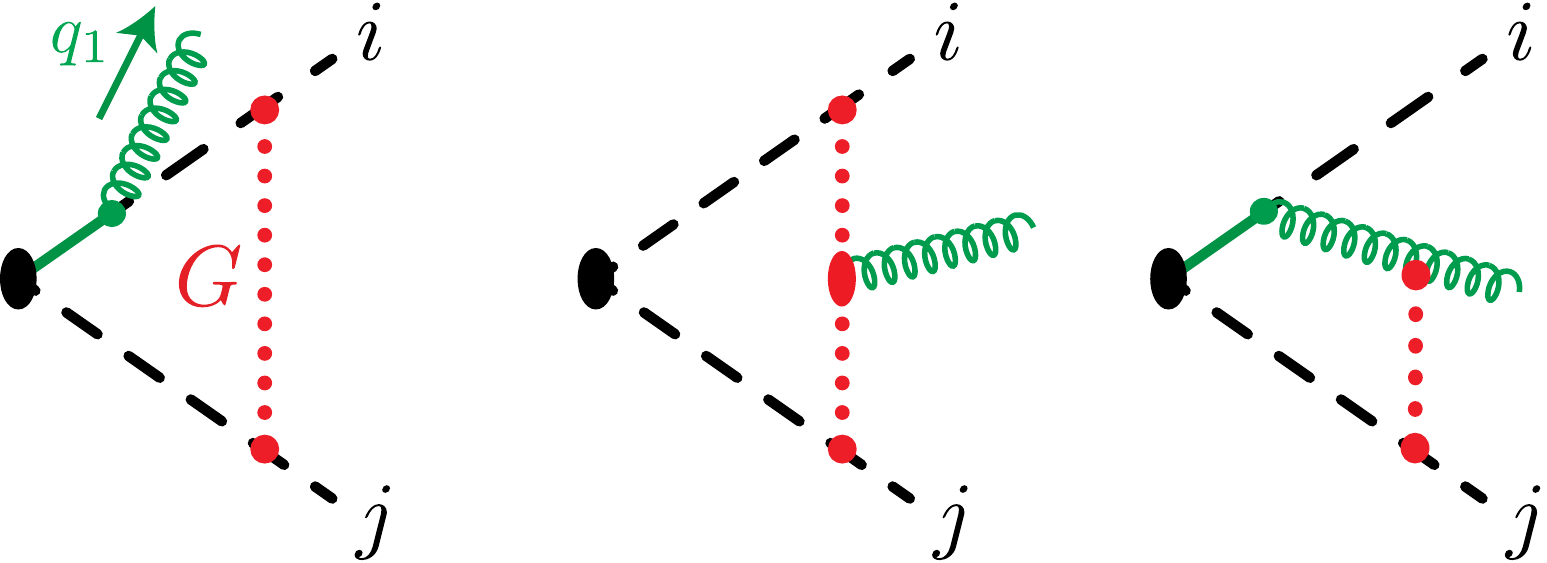}} \\ \vspace{5pt}
\subfigure[{Glauber exchange graphs in EFT$_{n+1}$.
\label{fig:all3} }] {\includegraphics[width=0.4\textwidth]{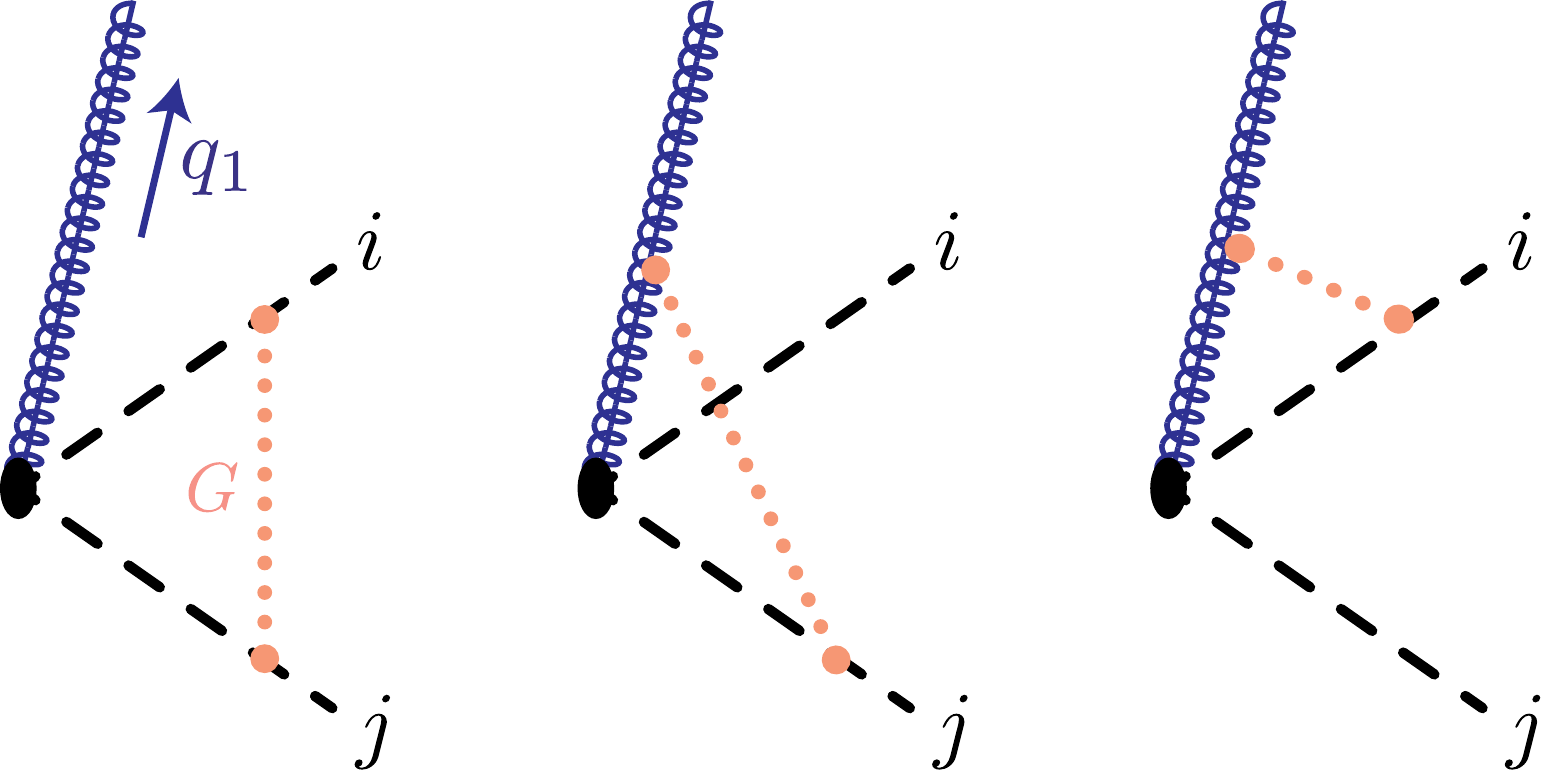}} \\ \vspace{5pt}
\subfigure[{An indicative selection of single soft gluon emission graphs with Glauber exchanges
in EFT$_{n+1}$. One must also consider other allowed permutations of gluon 2 and the Glauber vertices.
\label{fig:all4} }] {\includegraphics[width=0.4\textwidth]{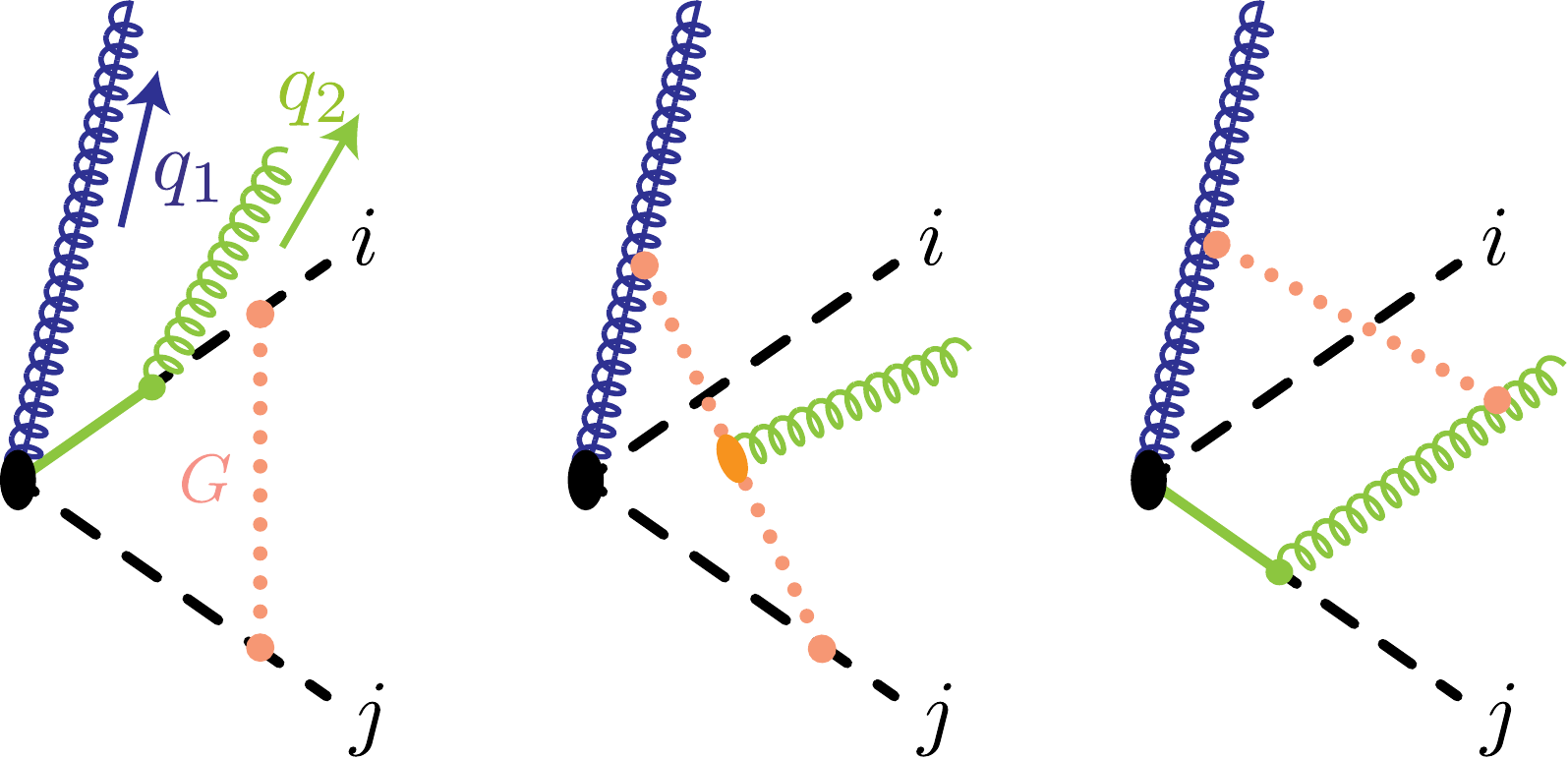}}
\caption{Diagrams needed to reproduce the imaginary part of one-loop and up to two real soft gluon emission amplitude. Labels $i$, $j$ correspond to hard particles. The gluon colored in blue is more energetic of the two ($q_{1}$). The green solid lines correspond to off-shell propagators. They have been depicted in this way to make the color ordering apparent. Graphs with Glauber exchange between a soft gluon and its parent are zero in Feynman gauge, and not shown.}
\label{fig:all}
\end{figure}

\subsection{Derivation from Glauber-SCET}
It is sufficient for us to consider the imaginary part of the one-loop result in \eq{AMFS}, which means we can focus solely on diagrams with a single Glauber exchange loop \cite{Rothstein:2016bsq}.
In overview, we use SCET to compute $ {\rm Im} \big | M^{(1)}_{2} \big ]$ from which the generalisation to $\big | M^{(1)}_{N} \big ]$ is reasonably simple. We exploit the hierarchy between the two soft emissions by defining two EFTs, EFT$_{n}$ and EFT$_{n+1}$. In both the EFTs the hard momenta ($p_1,\ldots p_n$) define $n$ collinear directions. In EFT$_{n+1}$ we treat the soft gluon $q_1$ as resolved by further restricting the fluctuations in its virtuality so that it may be considered as a new collinear mode. The virtuality of $q_2$ then sets the scale of soft fluctuations in EFT$_{n+1}$. The hard scattering operator in EFT$_{n+1}$, including $q_1$ as a collinear mode, is easily fixed based on symmetry considerations.

The operators necessarily come with Wilson coefficients which encode the UV physics, which we compute at one-loop. The Wilson coefficients carry dependence on the large momenta, $\omega_i = \bar n_i \cdot p_i = p_i^-$, in direction $n_i$ (with $\bar n_i$ being an auxiliary light like vector satisfying $\bn_i \cdot n_i = 2$). Crucially, we will see that the large momentum corresponding to the $q_1$ gluon, used in EFT$_{n+1}$, is transverse momentum $q_{1\perp}^{(ij)}$.

The hard scatterings in EFT$_{n}$ and EFT$_{n+1}$ are described by the operators,
\begin{align}\label{eq:OnSCET}
&O_n^{\rm hard \,scatter} = \\
&\int \Big(\prod_{i = 1}^n d\omega_i\Big) \big[O^{(0)}_{n} \big(\{\omega_i,n_i\}\big) \big| \prod_{i=1}^n \mb S_{n_i} \,
\big| {\cal C}_{n} \big( \{\omega_i \}, \mu \big)\big]\,
,\nn\\
&O_{n+1}^{\rm hard \,scatter} \nn
\\ & = \int \Big(\prod_{i = 1}^{n+1} d \omega_i \Big)
\big[O_{n+1}( \{ \omega_1 , n_1, \omega_i ,n_i\}) \big | {\cal C}_{n+1} ( \{\omega_1, \omega_i \},\mu) \big] \, .\nn
\end{align}
where
\begin{align}
&\big[O_{n+1}( \{\omega_1, n_1, \omega_i ,n_i\}) \big | \\&
= \big [O_n^{(0)}\big(\{\omega_i, n_i\}\big) \big | \Bigg[g\sum_{i = 1}^n \frac{n_i \cdot {\cal B}_{n_1\perp, \omega_1}^{(0)a}}{n_i \cdot q_1} \mb T_i^a\Bigg]\Big( \prod_{i = 1}^n \mb S_{n_i} \Big) \mb S_{q_1} \, ,
\nonumber
\end{align}
${\cal B}_{n_1\perp, \omega_1}^{(0)a}$ is the gauge invariant building block~\cite{Bauer:2001ct} for the harder of the two soft gluons (i.e. corresponding to $q_1$). The notation indicates that it is perpendicular to $n_1$. Here the operators and the Wilson coefficients are dual-vectors and vectors in the color space. Off-shell modes have been integrated out via a BPS field redefinition~\cite{Bauer_2002} of the collinear fields to obtain soft Wilson lines in directions $n_i$, and the adjoint Wilson line $\mb S_{q_1}$ in the direction $n_1$. Consequently, the operators $O^{(0)}_{n,n+1}$ consist solely of collinear fields. $\big | {\cal C}_{n} ( \{\omega_1 \},\mu) \big]$ and $\big | {\cal C}_{n+1} ( \{\omega_1, \omega_i \},\mu) \big]$ are the Wilson coefficients, they have a loop expansion for which we use the same notation as in \eq{JMFact1}.

For the general case of $n$ hard partons and $N$ soft emissions, the correspondence between QCD
and EFT$_{n}$ amplitudes is
\begin{align}\label{eq:QCD_EFT}
& \sum_{N} (g \mu^\eps)^{{N}} \mb J ({q_N}) \ldots \mb J ({q_1})
\big| {\cal M} (p_1 , \ldots, p_n) \big ] \nn
\\
& = \int \Big(\prod_{i = 1}^n d\omega_i\Big) \langle \{p_i\},\{q_j\} | \\ & \times {\rm T}\bigg\{O^{(0)}_{n} \big(\{\omega_i,n_i\}\big) \prod_{i=1}^n \mb S_{n_i} e^{\im \int d^4 x' O_G(x') } \bigg\}| 0 \rangle \big| {\cal C}_{n} \big( \{\omega_i \}\big)\big] \, ,
\nn
\end{align}
where $\{a_{i}\}$ is the set of colour indices for the hard particles and $\{C_{j}\}$ the colour indices for soft particles. Here, $O_G(x)$ consists of the Glauber operators from the SCET Glauber Lagrangian~\cite{Rothstein:2016bsq}. Note that
\begin{align}
\sum_{\{a_i\}} \Big[ \{a_i\} \Big| O^{(0)}_{n} \big(\{\omega_i,n_i\}\big) = \big [O_n^{(0)}\big(\{\omega_i, n_i\}\big) \big |.
\end{align}

By expanding the right hand side of \eq{QCD_EFT} to one-loop, we arrive at the result in \Ref{Angeles-Martinez:2016dph} using the following four steps, corresponding in turn to the four sets of diagrams in \fig{all}.
\begin{enumerate}
\item Expand \eq{QCD_EFT} perturbatively at tree level and one loop for $N = 0$ using EFT$_n$ (\fig{all1}). This fixes $|{\cal C}^{(0)}_n]$ and the imaginary part of $|{\cal C}^{(1)}_n]$.
\item Compute ${\rm Im} |M_1^{(1)}]$ by evaluating the matrix element for single-soft emission in EFT$_{n}$ (\fig{all2}) and combining it with $|{\cal C}_n]$ from step 1.
\item Use the result from step 2 to calculate $|{\cal C}^{(0)}_{n+1}]$ and ${\rm Im}|{\cal C}^{(1)}_{n+1}]$ by computing the one-loop, no real emission graphs in EFT$_{n+1}$ (\fig{all3}).
\item Calculate the one-loop single soft emission amplitude in EFT$_{n+1}$ (\fig{all4}) and combine it with ${\rm Im}|{\cal C}_{n+1}]$ from step 3 to arrive at ${\rm Im} |M_2^{(1)}]$.
\end{enumerate}
In evaluating these diagrams we made use of the SCET Feynman rules for soft emission from the Wilson lines $\mb S_{n_i}$, as well as those for the Glauber operators in $O_G(x)$, namely the collinear-Glauber, the Lipatov vertex and the soft-Glauber vertices provided in \Ref{Rothstein:2016bsq} appearing in \fig{all2}. The calculation applies to a generic hard scattering operator $O_n^{(0)}$.

Step 1 yields
\begin{align}
& \Big| {\cal C}_n^{(0)}(\{\omega_i\},\mu)\Big] = \big| M_0^{(0)}\big] \, , \label{eq:normC0} \\
&\Im \left |{\cal C}^{(1)}_n (\{\omega_i\},\mu) \right] = \sum_{i=1}^n\sum_{j<i} \overline{ \mb C}^{(ij)} (\mu, \sqrt{\omega_{ij}}) \big | M_0^{(0)}\big]\, , \label{eq:ImCn}
\end{align}
where
\begin{align} \label{eq:15}
&\overline{\mb C}^{(ij)} (m, \mu) \equiv g^2 \im (\mb T_i \cdot \mb T_j)
\frac{1}{2}
\int \frac{d^{d-2} \ell_\perp}{(2\pi)^{d-2}}
\bigg[\frac{\mu^{2\eps} }{\ell_\perp^2 -m^2}\bigg]\nn \\ &= (-\im \pi ) \frac{\alpha_s}{2\pi} (\mb T_i \cdot \mb T_j) \left(\frac{1}{\eps}+ \ln\frac{\mu^2}{m^2} +\mathcal{O}(\epsilon) \right) \, ,
\end{align}
where $m$ is a gluon mass used to regulate IR divergences, such that the EFT matrix elements are no longer scaleless. The $m$ dependence has canceled in arriving at \eq{ImCn}, such that the Wilson coefficient is IR finite. The EFT matrix elements, being UV divergent, are regulated in dimensional regularization, and hence depend on $\mu$. The EFT$_n$ is matched to the full theory at scales $\mu^2 \sim \{\omega_{ij}\}$. In IR and UV finite cases, where the integral in \eq{15} is bounded so that $\ell_{\perp} \in (a,b)$, we no longer need the gluon mass $m$ and
$\overline{\mb C}^{(ij)}$ becomes
\begin{align}
\mb C^{(ij)}(a,b) \equiv (-\im \pi ) \frac{\alpha_s}{2\pi} (\mb T_i \cdot \mb T_j) \ln\Big(\frac{b^2}{a^2}\Big).
\end{align}
The result of step 2 is that
\begin{widetext}
\begin{align}\label{eq:ImMn}
&{\rm Im}\Big| M^{(1)}_{1} \Big ] \equiv \Im \bigg (| \langle( q_1, \veps_1), \{p_i\} | O_n\big(\{\omega_i, n_i\}\big) \prod_{i=1}^n \mb S_{n_i} | 0 \rangle \Big | {\cal C}_n \big(\{\omega_i \} , \mu\big)\Big ]\bigg)^{(1)} =
\\
& g \sum_{i = 1}^n \Bigg(
\sum_{j<i}\left[\mb J^{(0)} (q_1) \mb C^{(ij)} (q_{1\perp}^{(ij)}, \sqrt{\omega_{ij}}) + \mb C^{(ij)}(m, q_{1\perp}^{(ij)}\big) \, \mb J^{(0)} (q_1) \right]
+ \sum_{j \neq i}
\mb C^{((n+1) j )}(m , q_{1\perp}^{(ij)}) \, \mb d^{(0)}_{ji} (q_1) \bigg]
\Bigg) \big|M_0^{(0)}\big], \, \nn
\end{align}
\end{widetext}
where $(\dots)^{(1)}$ indicates evaluating the bracketed expression at one loop. The graphs contributing to \eq{ImMn} are shown in \fig{all2}. The Lipatov vertex plays a crucial role in implementing the \textit{switch mechanism} identified in \Ref{Angeles-Martinez:2015rna}: the associated loop momentum is cut off in the IR by $q_{1\perp}^{(ij)}$ in order to account for production of $q_1$, i.e.
\begin{align}
&\int \frac{\df^{d-2} \ell_\perp}{(\ell_\perp^2 -m^2)}
-\frac{q_{1\perp}^2}{2} \int \frac{ \: \df^{d-2} \ell_\perp}{(\ell_\perp^2 -m^2) \,[ (\ell_\perp+ q_{1\perp})^2 -m^2]}
\nn \\
&\qquad = \int \frac{\df^{d-2}\ell_\perp}{( \ell_\perp^2 -m^2)}\Theta (| \vec \ell_\perp| - | \vec q_{1\perp} | ) \, ,
\end{align}
and the non-abelian structure of the vertex yields commutators of the form $[\mb J^{(0)}(q_1), \mb C^{(ij)}(q_{1\perp}^{(ij)}, \mu)]$. The commutators cancel terms from the hard scattering graph which are oppositely ordered in transverse momentum. The $\mu$ scale serves as a UV cutoff for the EFT matrix elements, and after combining it with the Wilson coefficients, is replaced by hard scales $\{\omega_{ij}\}$.
The third term results from the re-scattering graphs, which are UV finite on their own, and account for the non-trivial frame dependence in the third line in \eq{AMFS}.
\eq{ImMn} is in agreement with the calculation presented in \Ref{Angeles-Martinez:2015rna} (i.e. \eq{AMFS} for $N=1$).

After step 3:
\begin{widetext}
\begin{align}\label{eq:matchEFTn1}
\Big| {\cal C}_{n+1}^{(0)} ( \{\omega_1, \omega_i \},\mu) \Big] = & ~ \big| M_0^{(0)}\big] \\
\mb J^{(0)} (q_1) \Im \left|{\cal C}^{(1)}_{n+1} ( \{q_{1\perp}^{(ij)},\omega_i\},\mu) \right] = & ~ \mb J^{(0)}(q_1) \sum_{i=1}^n\sum_{j<i} \mb C^{(ij)}(q_{1\perp}^{(ij)},\sqrt{\omega_{ij}}) \big|M_0^{(0)}\big] \nn \\
&+ \sum_{i = 1}^n \bigg( \sum_{j<i} \overline{\mb C}^{(ij)}(\mu, q_{1\perp}^{(ij)}) \mb J^{(0)} (q_1)
+ \sum_{j \neq i} \overline{\mb C}^{((n+1) i)}( \mu, q_{1\perp}^{(ij)}) \mb d^{(0)}_{ij} (q_1)
\bigg) \big|M_0^{(0)}\big] \nn \, .
\end{align}
\end{widetext}
The structure of Wilson coefficient in \eq{matchEFTn1} is analogous to that in \eq{ImMn} with $m \ra \mu$. This is because the matrix elements in EFT$_{n+1}$ ought to account for all the IR divergences which cancel in the matching calculation. Each term in \eq{matchEFTn1} has a physical interpretation: the first term in the second line \eq{matchEFTn1} has the same functional form as $\mb J^{(0)}(q_1){\rm Im}\big| {\cal C}^{(1)}_{n} \big]$ (see \eq{ImCn}), encoding the same UV physics; whereas terms in the third line are new, where it is the quantity $\omega_1 = q_{1\perp}^{(ij)}$ that sets the hard scale for EFT$_{n+1}$.

Finally, step 4 yields
\begin{widetext}
\begin{align}\label{eq:AMFS_SCET}
& {\rm Im} \Big| M^{(1)}_{\textcolor{Green}{2}} \Big ] \equiv \Im
\bigg( \langle( \textcolor{Green}{q_2}, \veps_2) ,( \textcolor{Green}{q_1}, \veps_1) ,\{p_i\} | O_{n+1}( \{ \textcolor{Green}{q^{(\textcolor{Red}{ij})}_{1\perp}}, n_1, \textcolor{Green}{\omega_i} ,n_i\}) | 0 \rangle \Big | {\cal C}_{n+1} (\{ \textcolor{Green}{q^{(\textcolor{Red}{ij})}_{1\perp}}, \omega_i \},\mu) \Big] \bigg)^{(1)}\nn \\
&=
g^2\, \Bigg[ \mb J^{(0)}(\textcolor{Green}{q_2}) \mb J^{(0)}(\textcolor{Green}{q_1})\sum_{\textcolor{Red}{i} = 1}^n \sum_{\textcolor{Red}{j} \neq \textcolor{Red}{i}}
\mb C^{(\textcolor{Red}{ij})}(\textcolor{Green}{q^{(\textcolor{Red}{ij})}_{1\perp}},\textcolor{Red}{\sqrt{\omega_{ij}}}) \\
&\qquad \qquad + \mb J^{(0)}(\textcolor{Green}{q_2}) \sum_{\textcolor{Red}{i} = 1}^{n}
\bigg(
\sum_{\textcolor{Red}{j} < \textcolor{Red}{i}}
\mb C^{(\textcolor{Red}{ij})} ( \textcolor{Green}{q^{(\textcolor{Red}{ij})}_{2\perp}}, \textcolor{Blue}{q^{(\textcolor{Red}{ij})}_{1\perp}} )
\mb J^{(0)} (\textcolor{Blue}{q_1})
+ \sum_{\textcolor{Red}{j}\neq \textcolor{Red}{i}} \mb C^{(\textcolor{Blue}{(n+1)\textcolor{Red}{i}})} (\textcolor{Green}{q_{2\perp}^{(\textcolor{Blue}{(n+1)i})}}, \textcolor{Blue}{q^{(\textcolor{Red}{ij})}_{1\perp}}) \, \mb d^{(0)}_{\textcolor{Red}{ij}}(\textcolor{Blue}{q_1})
\bigg)
\nn \\
&\qquad \qquad
+
\sum_{\textcolor{Red}{i} = 1}^{n+\textcolor{Green}{1}} \bigg( \sum_{\textcolor{Red}{j} < \textcolor{Red}{i}} \mb C^{(\textcolor{Red}{ij})}(m, \textcolor{Blue}{q^{(\textcolor{Red}{ij})}_{2\perp}}) \mb J^{(0)}(\textcolor{Blue}{q_2})
+ \sum_{\textcolor{Red}{j}\neq \textcolor{Red}{i}} \mb C^{(\textcolor{Blue}{(n+2) \textcolor{Red}{i}})}(m, \textcolor{blue}{q^{(\textcolor{Red}{ij})}_{2\perp}})\, \mb d^{(0)}_{\textcolor{Red}{ij}}(\textcolor{blue}{q_2}) \bigg) \mb J^{(0)} (\textcolor{blue}{q_1})\Bigg] \Big| M_0^{(0)}\Big]\, .\nn
\end{align}
\end{widetext}
This equation is exactly equal to the imaginary part of \eq{AMFS} when $N=2$. It is straightforward to extend the result above to the complete result in \eq{AMFS} by including more tree-level soft gluons.

In conclusion, using SCET we have presented a very compact derivation of the result in \Ref{Angeles-Martinez:2016dph} that involved only a handful of diagrams, and where each of the diagrams has a clear physical relevance. The simplicity of this derivation makes the extension to higher orders forseeable: for example, the two-loop extension will only require the effective one-loop collinear-Glauber, soft-Glauber and one-loop Lipatov vertices. We leave this to future work.

This work is supported by the UK Science and Technology Facilities Council (STFC) under grant number ST/T001038/1.
\bibliography{qcd}

\end{document}